# Trustworthy Smart Band:
# Security Requirement Analysis with Threat Modeling

Suin Kang, Hye Min Kim and Huy Kang Kim

Graduate School of Information Security, Korea University

**Abstract.** As smart bands make life more convenient and provide a positive life-style, many people are now using them. Since smart bands deal with private information, security design and implementation for smart band system become necessary. To make a trustworthy smart band, we must derive the security requirements of the system first, and then design the system satisfying the security requirements. In this paper, we apply threat modeling techniques such as Data Flow Diagram, STRIDE, and Attack Tree to the smart band system to identify threats and derive security requirements accordingly. Through threat modeling, we found the vulnerabilities of the smart band system and successfully exploited smart bands with them. To defend against these threats, we propose security measures and verify that they are secure by using Scyther which is a tool for automatic verification of security protocol.

## 1    Introduction

Smart band automatically measures the number of steps, heart rate, and sleep time, helping a user to have a positive lifestyle. Also, It is connected to a smartphone and performs various functions such as alerting the user when the smartphone receives messages, monitoring the current user state, setting the alarm, etc. As these functions have a direct impact on daily life, the smart band system requires a high level of trustworthiness. If an attack becomes successful, then sensitive health information and private information can be leaked (e.g., moving path, daily life pattern, etc.). Also, the attacker can obtain control of the smart band and use it freely.

Many studies have uncovered attacks and vulnerabilities of smart bands. Zhou *et al*. analyzed threats of a commercial fitness tracker and found that it transmits the login information and data in plaintext without encryption [1]. Lee *et al*. succeeded in detecting weakness of wearable service and set up attack scenarios [2]. Goyal *et al*. obtained private data of wearable health tracker and conducted DoS attack successfully [3]. Seneviratne *et al.* analyzed security threats of wearable devices based on three categories: Confidentiality, Integrity, and Availability [4]. These studies show that many devices have been released with vulnerabilities. If smart bands are commercialized with unknown vulnerabilities, the user's privacy and sensitive information can not be protected. Therefore, in order to construct a trustworthy smart band system, it should be able to respond to unknown vulnerabilities as well as known ones and clearly identify attack points.

The trustworthy system refers to a system that can operate safely in any situation considering availability, reliability, security, and safety [5, 6]. A series of processes for developing and operating the trustworthy system is called information assurance. Achieving information assurance requires the step-by-step assurance from the security requirements analysis and design to the implementation and operation of the system. The most important part in achieving information assurance is to derive security requirements because the whole system is designed and implemented taking into account the security requirements of the system.

In this paper, we apply threat modeling techniques such as Data Flow Diagram [7, 8], STRIDE [9, 10], and Attack Tree [11] to the smart band system to identify security threats and derive security requirements accordingly. With this process, it can respond to unknown vulnerabilities as well as known ones and identify attack points in advance. For example, we could find a vulnerability in the connection process for the smart bands and successfully gain system privilege by using the vulnerability found by threat modeling. To deal with the connection problem and other vulnerabilities, we provide secure connection and communication protocols. Also, we present a simple and practical (considering the smart band environment) key exchange protocol for encryption in secure connection process. Finally, we verify the security measures using an automatic verification tool. An overview of the proposed approach is shown in **Fig. 1**.

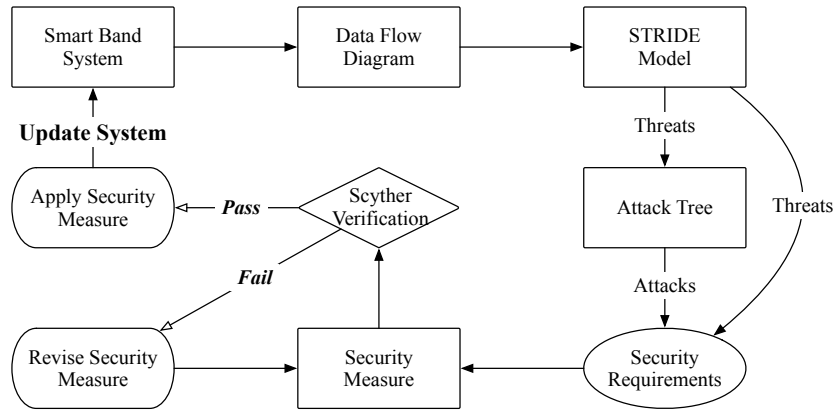

**Fig. 1.** Overview of the proposed approach.

**Contributions.** The contributions of this paper are as follows:

– We derive vulnerabilities and security requirements for the smart band system. These are useful for the service providers to construct the trustworthy system achieving information assurance.

— We demonstrate that threat modeling analysis is effective in deriving threats and attack scenarios for smart bands with a real experiment. We found a vulnerability on smart band connection process. Using this vulnerability, an attacker can spoof an authorized user to take control of the smart band and use it like the normal user.

— We propose security measures to solve the problem on the connection process and protect the smart band system. Considering the hardware characteristics of smart bands, the measures are designed to be simple and easy to apply to the system. Also, we validate the measures using Scyther [12] which is an automatic verification tool.

## 2    Preliminaries

### 2.1    Security of smart band

Zhou *et al.* analyzed the threat of the commercial fitness tracker [1]. The fitness tracker, as called Fitbit, transmits the login information and HTTP data in plaintext without encryption when communicating with the web server. This allows attackers to access data easily without authentication. To handle this vulnerability, Rahman *et al.* proposed FitLock methodology [13]. However, a risky attack is still possible even if FitLock is applied to the system. Therefore, for the fitness trackers that are connected to the online networks, it is necessary to analyze the vulnerabilities more closely and to design security measures against them. Lee *et al.* analyzed the system to find the vulnerability of wearable service with the perspective of devices, gateways and server [2]. They set up three attack scenarios: Pairing to unauthorized devices, information leakage from fake gateways, and malicious code injection. After that, those scenarios were applied to a commercial smart band, and they proved that the attacks were successful. Finally, a secure authentication procedure is proposed to prevent those attack scenarios. Goyal *et al.* analyzed wearable health trackers, focusing on the security and privacy issue [3]. They showed how these devices are vulnerable to several attacks and formulated an assessment table after security analysis. They used *GattTool* for gaining access to user's private data and simulated DoS attack with the commercialized devices. Also, they modified the source code of the mobile application to get the user's data. Furthermore, they found HTTP and SSL vulnerabilities in the server certificate validation. Fereidooni *et al.* analyzed 17 wearable devices and revealed severe vulnerabilities which can be exploited easily [14]. They demonstrated that they could get the user's private data and modify the data of the device directly by spoofing the authenticated user. They concluded that End-to-End encryption should be implemented and a digital signature must be used in the fitness tracker's communication protocol.

According to these researches, many of the devices were released with errors while they were not aware of the vulnerability. If a smart band with the unknown vulnerability is commercialized, the users that purchased the device are not protected against the threats. Therefore, it is necessary to progress threat analysis with threat modeling to detect not only unknown vulnerabilities but also known ones in advance. The provider of the smart band must apply security measures to make the system secure and trustworthy.

## 2.2    Data Flow Diagrams (DFD)

To find out and reveal attack surfaces and attack vectors explicitly, we need to understand the whole system of the smart band. With Data Flow Diagrams, we can abstract the system structure and see the data flow of the system at a glance. Because DFD describes the data flows of the components in a system, it is ideal for threat modeling [7]. Through the abstraction process, it can remove parts that are not necessary to be analyzed and make it easy to understand the whole structure of the system and the flow of data [8]. Because attacks usually occur as data flows, especially when the data crosses the trust boundary, we need to clarify what happens as data flows cross trust boundaries. Therefore, security experts must draw DFD that describes the flow of data clearly to identify the vulnerabilities in advance. Also, since DFD is drawn on an elemental basis, it is easy to find element-level vulnerabilities. The elements of DFD are described in the below.

- **External entity (Rectangle).** The external entity is any entity outside the system that interacts with inner elements. It generates the input and takes the output from a process.
- **Process (Circle).** The process is a task that performs functions to handle data within the application. It takes inputs and generates outputs for the function. This is necessary when exchanging data between different elements.
- **Data flow (Directed arrow).** The data flow represents the data transmission between other elements. The direction of data flow can be represented by the arrow.
- **Data store (Two parallel lines).** The data store is used for storing data in the system. It only stores and transmits the data, but it cannot perform functions with the data.
- **Trust boundary (Red dashed lines).** The trust boundary describes the borderline where the trust level changes. This makes it easy to identify where the privilege changes and where the data comes from the untrusted sources.

## 2.3    STRIDE

STRIDE is a classification scheme developed by Microsoft for grouping threats into categories [9, 10]. It is derived from an acronym for the following six threat categories, Spoofing, Tampering, Repudiation, Information disclosure, Denial of service, and Elevation of privilege. The detailed explanation of STRIDE is as follows:.

- **Spoofing.** An attacker pretends to be an authorized user so that the attacker can get access to the target of the whole system and get user's authentication information.
- **Tampering.** An attacker modifies data to deceive the user. The target data is usually in a data store or data flow between entities.
- **Repudiation.** An attacker denies performing an action, such as getting or sending data and it makes it difficult to identify who does malicious actions.
- **Information disclosure.** The information is exposed to an attacker unauthorized to access it. For dealing with this problem, it is necessary to prevent tampering the privilege and care about the point where data leaks.

- **Denial of service.** An attacker interrupts the user to get the resource of services. To prevent it, availability and reliability of the system should be improved.
- **Elevation of privilege.** An attacker gains the privilege without proper authorization process. After an attack is completed successfully, an attacker can gain access and penetrate to the system.

Using DFD together, we can easily identify the attack points of the target system. STRIDE can be applied to various fields because there is no restriction on the analysis system and it consists of standardized analysis processes. Dev *et al*. analyzed the vulnerability of the Google Chrome API using STRIDE, showing the possibility of various threats [15]. They provided security guidelines to developers and informed developers of Chrome-extensions vulnerabilities that can occur depending on the functionality of the APIs. Karahasanovic *et al*. analyzed the vulnerability of automobiles adopted IoT technology using both STRIDE and TARA techniques in accordance with AUTOSAR standards of automobiles [16]. They designed the template of threat modeling to contrive a security mechanism that will be useful for the autonomous mobile industry. Cagnazzo *et al*. analyzed the threat of mobile health systems that are connected with IoT devices, using STRIDE threat model [17]. They especially focused on the network, such as BAN, WLAN. They abstracted the mobile health system with DFD. After threat modeling, they proposed possible mitigation strategies that apply encryption techniques and authentication to the system.

Using STRIDE is helpful to find the vulnerability of the target easily. Therefore, we found out vulnerabilities of smart bands using STRIDE and proposed security measures against them in this paper.

### 2.4 Scyther

Scyther is an open source automatic tool that demonstrates the security of the protocols used on the Internet and open networks. Compared with other automatic analysis tools, the analysis is possible without using abstraction technology, and the speed of verification is faster than others. Scyther is also widely used for educational purposes to better understand the protocol configuration, and it was used in research papers to verify the safety of protocol [12]. Basin *et al*. analyzed the vulnerability using Scyther and suggested security measures to prevent it in the ISO/IEC 9798 standard protocol [18]. After this, ISO updated the standard protocol and distributed it. Cremers, C. proceeded to formalize the Internet key exchange protocol and analyzed the phase of it so that they found an unknown vulnerability of the protocol using Scyther [19]. Also, they provided the most comprehensive view of the security properties for IKE protocols.

## 3 Threat Modeling

In this section, we analyze the vulnerabilities and threats of the smart band system. Threat analysis is performed for basic services provided by the smart band including notification, alarm, data transmission. The threat analysis process is conducted through security threat modeling, and the steps are as follow:

1. After defining the boundary and scope of the system, we identify important assets. At this time, DFD is drawn to clarify asset identification and threat analysis.
2. Based on the DFD, we apply STRIDE model to the system to derive possible threats.
3. The threats derived from STRIDE model could be used to attack the smart band system. By drawing attack tree, we can learn how the threats can be used in attacks.

### 3.1 Data Flow Diagram of Smart Band System

We draw DFD to analyze the threats of the smart band system and identify data flows. Because the attacker usually attacks when the target data crosses the trust boundary, it is easy to identify the security threats if DFD is correctly drawn.

As shown in the DFD from **Fig. 2** the smartphone, the smart band, and the web server exchange the user information. If attackers can get this information, they can violate privacy and use the information for a malicious purpose. Therefore, we define the main assets of the smart band system as the user information and the smart band service itself. The system is analyzed focusing on the user information and the smart band service. The smart band system could be divided into three parts and the description of each part is following:

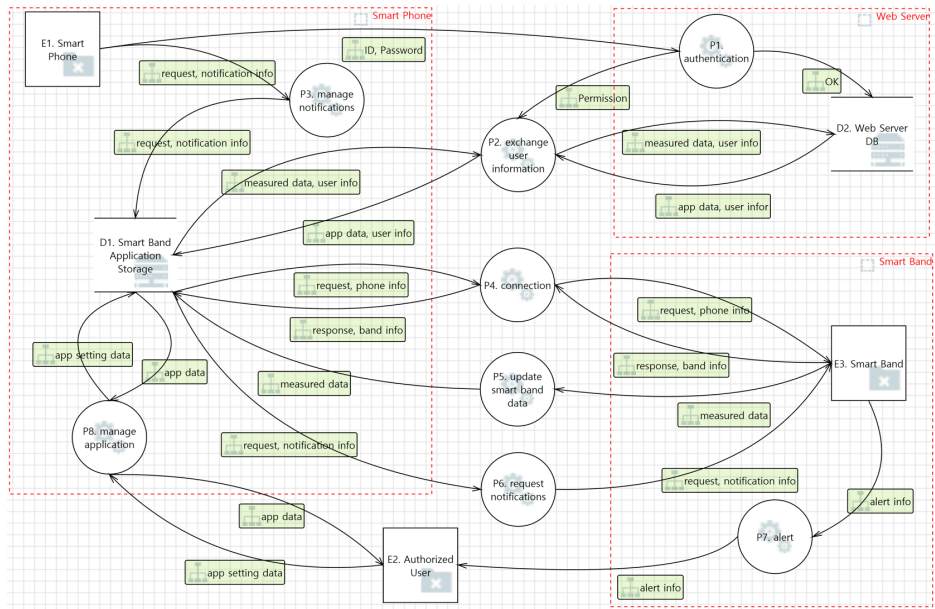

**Fig. 2.** Data Flow Diagram of smart band system

— The upper part of **Fig. 2** shows the process of exchanging data between the web server and the user's smartphone. The ID and password entered in the smartphone (E1) are sent to the web server to authenticate the user (P1) and then transmitted to the web server database (D2). The web server transmits the user information of the corresponding ID and the application data (P2) to the smart band application store (D1). Also, the application sends the measured data, such as the number of steps and the heart rate, to the web server database.

— The center of **Fig. 2** shows the process of connection and exchanging data between the user's smartphone and the smart band. When the smartphone receives a message, it checks the type of the message and transmits notification information to the application (P3). The smart band application requests the connection and sends the smartphone information to the smart band to connect with the smart band (E3). The smart band establishes a connection by transmitting a response and band information (P4). When the connection is completed, the smart band goes through a synchronization process and updates the measured data to the application (P5). After that, the application transmits notification information to the smart band along with the notification request (P6).

— The lower part of **Fig. 2** shows the process of exchanging data between the authorized user (E2) and the application, and alerting the user. The smart band transmits the notification information to the authorized user (P7) by vibrating and displaying icons or numbers. The authorized user sends application setting information, such as timer alarm, to the application and gets the smart band data from the application (P8).

### 3.2    STRIDE Threat Analysis

In the above, we defined the main asset of the smart band system as the user information and the service itself. Considering the main assets of the system, we set the attacker's goal to obtain the user information and conduct denial of service attack. We have applied the STRIDE model to identify the possible threats of smart band service. For applying STRIDE model to the smart band system, we used Microsoft Threat Modeling tool. Microsoft Threat Modeling tool [20], [21] generates the report of threat modeling automatically based on the DFD. It automatically derives 147 threats. Considering the previous researches on smart bands security [1]-[4], [13], [14], [17], [22], we select 33 important threats that could be applied to achieve the attacker's goal and enumerate them in **Table 1**.

**Table 1.** STRIDE threat analysis of smart band system

| Element | Type | Threat Description |
|---------|------|--------------------|
| E1. Smartphone | S | T1. Attacker spoofs smartphone to get user information. |
|  | R | T2. Attacker sends data and denies this later. |
| E2. Authorized User | S | T3. Attacker spoofs an authorized user to manipulate application data. |
|  | R | T4. Attacker sets application data and denies this later. |
| P1. | S | T5. Attacker spoofs the smartphone or web server to obtain ID/ PW of user |

| Authentication | | or access permission for information exchange. |
|---|---|---|
| | T | T6. Attacker modifies ID/ PW of user to disturb authentication. |
| | I | T7. Attacker obtains ID/ PW. |
| | D | T8. Attacker makes it impossible to perform the normal authentication through excessive authentication requests. |
| | E | T9. Attacker gives the higher privilege to someone who does not have it. |
| P2. Exchange User Infor-mation | S | T10. Attacker spoofs the authorized user's smartphone to transmit wrong data or gain access to the transmitted data. |
| | T | T11. Attacker modifies application data or user information and transmits it. |
| | I | T12. Attacker obtains user information or application data. |
| | D | T13. Attacker makes it impossible to exchange data through excessive data transmission. |
| | E | T14. Attacker gives the privilege to someone who could not access to application information and user information. |
| P4. Connection | S | T15. Attacker pretends to be an authorized user to control smart band. |
| | T | T16. Attacker modifies the smartphone/ smart band information and transmits them. |
| | I | T17. Attacker obtains smartphone/ smart band information and connection information. |
| | D | T18. Attacker makes it impossible to connect through an excessive connection request. |
| | E | T19. Attacker gives the privilege to someone who could not access to connection information. |
| P6. Request Notifications | S | T20. Attacker spoofs the authorized user's smartphone to transmit wrong notification information. |
| | T | T21. Attacker modifies notification information and transmits it. |
| | I | T22. Attacker obtains notification information. |
| | D | T23. Attacker makes it impossible to transmit the normal notification request through excessive notification requests. |
| | E | T24. Attacker gives the privilege to someone who could not access notification request. |
| D1. Smart Band Application Storage | T | T25. Attacker modifies the data in the storage and inserts wrong data. |
| | I | T26. Attacker accesses the datastore and obtains application information and user information in the storage. |
| | D | T27. Attacker inserts data too much so that the storage does not work |
| | | properly. |
| D2. Web Server Database | T | T28. Attacker modifies the data in the web server database and inserts wrong data. |
| | I | T29. Attacker accesses the database and obtains application information and user information in the web server database. |
| | D | T30. Attacker transmits data excessively to disturb normal data exchange. |
| P1 → D2 | T | T31. Attacker modifies authentication result so that the authorized user could not access web server database, or an unauthorized user could access web server database. |
| | I | T32. Attacker obtains authentication result. |
| | D | T33. Attacker transmits authentication data excessively to disturb the nor- mal authentication. |

### 3.3    Attack Tree

Attack tree is a tool to arrange possible attacks on the system hierarchically, and so it is helpful to find attack scenario of the system [11]. We construct attack trees to find possible attack scenarios on the smart band system by analyzing the association be- tween the threats derived from STRIDE model and the attacker's goal. When designing the attack tree, we did not consider the side-channel attack, social engineering attack, and outside of the smart band system boundary. We refer to research papers previously studied on smart bands and wearable devices security [1]-[4], [13, 14], [17], [22] and results of STRIDE threat analysis for the reasonable attack trees.

The root nodes of the attack trees are the attacker's ultimate goals, and the leaf nodes are attacks. To achieve the ultimate goals, the attackers can use one of the leaf nodes. As we descend along the tree, we can come up with an attack scenario. We created two trees based on two targets of the attacker and described for each tree. The attack trees are shown in **Fig. 3**.

**Smart Band Denial of Service.** Smart band denial of service could be achieved by accomplishing one of the following goals: `Transmit Malicious Information to User`, `Excessive Data Transmission`, and `Taking over the Smart Band Control`.

1. To transmit malicious information to the user, an attacker saves previously transmit- ted information and retransmits it later(A1). Also, the attacker modifies data and inserts it(A2). For these attack, spoofing an authorized user or getting the higher privilege could be used to access and modify the information. By packet sniffing during data transmission, an attacker could save the packets and retransmit the same packets.

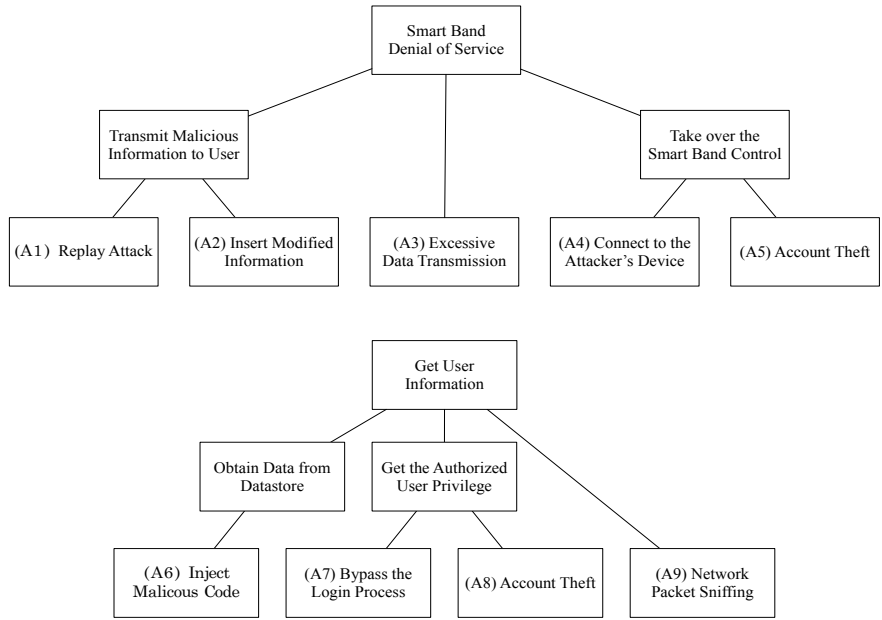

**Fig. 3.** Attack tree for smart band system

2. In the case of excessive data transmission(A3), it is possible to interrupt the use of the smart band service by transmitting excessive data and making the smart band vibrate continuously. To perform the attack effectively, attackers could target the user authentication process or the process of exchanging user information.

3. Attacks that prevent the use of smart band services by taking over the smart band control are achieved by connecting the smart band to the device of the attacker(A4) and stealing account of the normal user(A5). An attacker can connect to a smart band by pretending to be an authorized user during the connection process. Also, an attacker can acquire the user information during the authentication process to perform the account theft attack.

**Get User Information.** Getting user information could be achieved by accomplishing one of the following goals: `Obtain Data from Datastore', `Get the Authorized User Privilege', and `Network Traffic Sniffing'.

1. To obtain data from datastore, an attacker gets data from user's phone or smart band by using file recovery tools or injecting malicious codes in them(A6). These attacks can occur with an attacker having access to the datastore. Alternatively, an attacker can spoof as a normal user or elevate the privilege to do unauthorized conducts.

2. The user information can be obtained by acquiring the authorized user privilege. This can be done by bypassing the login procedure(A7) and stealing the authorized user account(A8). For these attacks, an attacker can target the authentication process or the user information exchange process.

3. An attacker obtains information through network traffic sniffing when the data is transmitted [22]. An attacker can obtain the user information by sniffing the transmitted data(A9) during the communication between the smartphone and the web server or during the communication between the smartphone and the smart band.

In addition to the threats mentioned above, attackers can attempt to combine various threats derived from STRIDE threat analysis. Based on the threats and attack scenarios derived from threat modeling, we exploited vulnerabilities in the connection process of the smart bands and successfully gained system privilege. Our attack demo video can be found on the following website: https://youtu.be/QFb1AV7yUas [23]. If the smart band and the normal user's smartphone are already connected, the attacker could not easily connect the smart band. However, if they are not connected, the attacker could easily gain access to the smart band (see **Fig. 4**). This attack is possible because the smart band did not check whether the smartphone is an authorized device or not when connecting the smart band to the smartphone. If the user left the smart band, an attacker can acquire the smart band physically and connect it to the attacker's device without the user authentication process. Using this vulnerability, the attacker can spoof an authorized user to take control of the smart band and use the smart band like the normal user. Therefore, user authentication process is necessary for secure connection and we propose it in 4.2. This experiment confirms that our threat modeling method is effective in detecting system vulnerabilities.

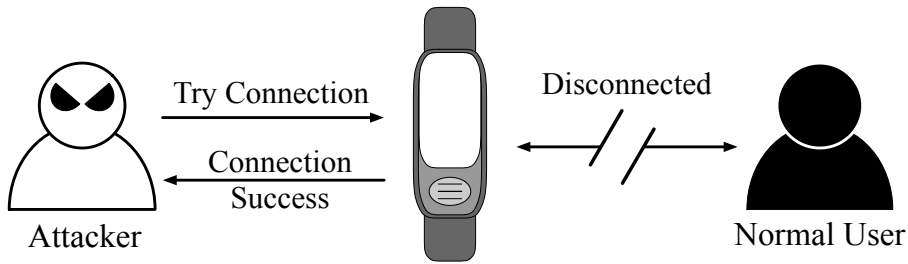

**Fig. 4.** Attack scenarios using vulnerabilities on the connection process.

## 4 Security Requirements and Security Measures

In this section, we derive the security requirements of the smart band system and provide security measures satisfying the security requirements and validate the security measures.

### 4.1 Security Requirements

As mentioned above, the attacker's ultimate goals are `Smart Band Denials of Service' and `Get User Information'. Since the security requirement is intended to prevent the attackers from achieving their goals when we derive security requirements, we consider the attack scenarios obtained from the attack tree.

- **Adding timestamp in the packet**. To prevent the replay attack(A1), timestamp should be added in the packet for all data exchange process and connection process.
- **Authentication**. Even if an attacker modifies a message, message authentication process can detect it and prevent inserting modified information(A2). Also, the user authentication process should be used to prevent connecting to an attacker's device (A4), stealing account(A5, A8), and bypassing the login process(A7).
- **Encryption**. Encryption makes it difficult for an attacker to read the original message and modify it. Therefore, packet encryption process should be used for preventing message modification(A2) and network packet sniffing(A9). Also, all sensitive data in the database must be encrypted before saving it. Then, even if the attacker gets the physical storage, the attacker cannot read the data(A6).
- **Traffic analysis and intrusion detection**. It is difficult to block the excessive data transmission(A3) entirely. Therefore we need to mitigate it through traffic analysis and intrusion detection on the server-side.
- **Secure account management**. All the users must manage their accounts securely to prevent stealing account(A5, A8). The service provider should keep informing the user of the risk of account theft and manage the datastore securely.

### 4.2 Security Measures

Looking at the smart band DFD and security requirements, two major parts require authentication and encryption when transmitting a packet. The first part is the communication between the smartphone and the web server. When a packet comes and goes between a smartphone and a web server, the packet must have authentication information for the user and be encrypted. The second part is the communication between the smartphone and smart band. In this case, a secure connection is necessary to solve the problem of the connection process mentioned above. For the secure connection, encryption and the key generation process for encryption are essential. In this subsection, we present security protocols and a simple and practical (considering the smart band environment) key exchange protocol.

**Communication between Smartphone and Web Server.** The communication between the smartphone and the web server is divided into two phases. The first step is the login phase. To communicate with the web server, the user first logs in and sends data to the web server. In this process, authentication and data encryption procedures are needed to prevent attacks such as replay attack, login bypass, and account theft. The second step is the data transmission phase. In this process, the symmetric key encryption algorithm is used because the data size may be large. In exchange of the symmetric key, we use the station-to-station protocol [24]. Another secure protocol could be used for the key exchange instead. After key exchange, we should encrypt the timestamp, each identification information, and data for communication. It is assumed that the certificates of smartphone and web server are exchanged through secure channels. The formal security measure is shown in **Fig. 5** The procedure is as follows:

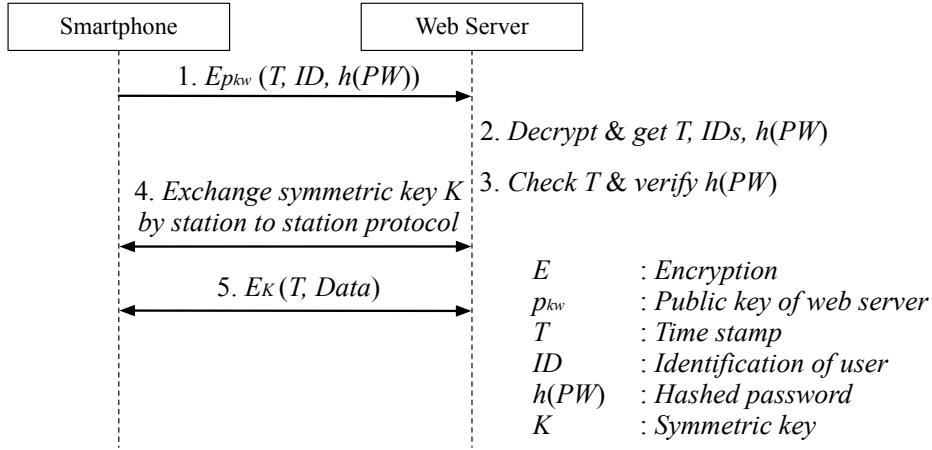

**Fig. 5.** Smartphone and web server communication protocol

1. The smartphone encrypts timestamp, the user ID, and hashed password using the public key of the web server.
2. The web server decrypts the encrypted contents using the private key of the web server to obtain the timestamp, the smartphone user ID, and the hashed password.
3. The web server verifies the validity of the timestamp and checks the corresponding ID and hashed password. If it passes normally, the web server approves the login.
4. Obtain a symmetric key for communication between the smartphone and the web server using the station-to-station protocol.
5. The smartphone and the web server use the symmetric key to encrypt the timestamp, the identification information, and the data.

**Connection and Communication between Smartphone and Smart Band.** The secure connection between smartphone and smart band is necessary so that an attacker cannot connect the smart band to the attacker's device. To establish a secure connection, the smart band must store the smartphone information of the authorized user and carry out the user authentication during the connection process and data transmission process. When transmitting the user smartphone information to the smart band, it must be encrypted. For encryption, it is necessary to exchange a symmetric key. We propose a simple symmetric key generation process for encryption (see **Fig. 6**). U1, U2, B1, B2, B3, P1, P2, and P3 denote the order of action of the user (U), the smart band (B), and the smartphone (P) respectively. The random number generator is assumed to be secure because it is out of the scope of this paper. The process is as follows:

1. The smartphone requests a connection to the smart band (P1).
2. The smart band that receives the connection request transmits a response message and displays a number on the screen to inform the user of a specific number (B1).
3. The smartphone receives a response message and changes the screen for the user to input the number (P2).

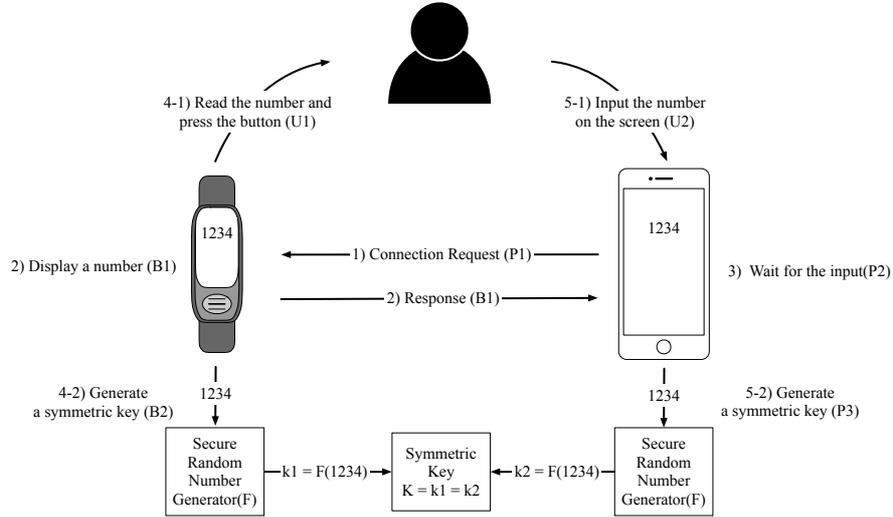

4-1) Read the number and press the button (U1)

2) Display a number (B1)

1) Connection Request (P1)

2) Response (B1)

5-1) Input the number on the screen (U2)

3) Wait for the input(P2)

4-2) Generate a symmetric key (B2)

5-2) Generate a symmetric key (P3)

Secure Random Number Generator(F)

k1 = F(1234)

Symmetric Key K = k1 = k2

k2 = F(1234)

Secure Random Number Generator(F)

**Fig. 6.** Symmetric key generation process for secure connection

4. The user reads the number on the smart band and presses a button (U1). Then, the smart band generates a symmetric key (B2).
5. The user inputs the number into the smartphone (U2). Then, the smartphone generates a symmetric key (P3).

Since this key generation process is used for the initial connection setting and the user reads the number and inputs it into the smartphone directly, it is reasonable to assume that the symmetric key generation is secure. After exchanging the symmetric key, the smartphone encrypts the connection request message and timestamp with the symmetric key and transmits it. When the smart band receives and decrypts the message, If the message format is wrong, then the smart band ignores it. After the connection setup is completed, the smartphone and the smart band encrypt timestamp and data and send the encrypted contents whenever they want to communicate. If the user wants to connect the smart band to a new smartphone, the old smartphone transmits the disconnection request in the data field to release the smartphone connection. The formal security measure for connection and communication between smartphone and smart band is shown in **Fig. 7**. It is assumed that the symmetric key exchange process has been executed in advance through the method mentioned earlier. The procedure is as follows:

1. The smartphone encrypts the timestamp and the connection request message with the symmetric key and sends the encrypted contents.
2. The smart band decrypts the message and checks the message format. If the format is wrong, the smart band ignores the message.
3. The smart band encrypts the timestamp and the connection response message with the symmetric key and sends the encrypted contents.
4. For communication, encrypts the timestamp and data with the symmetric key and transmits the encrypted contents.

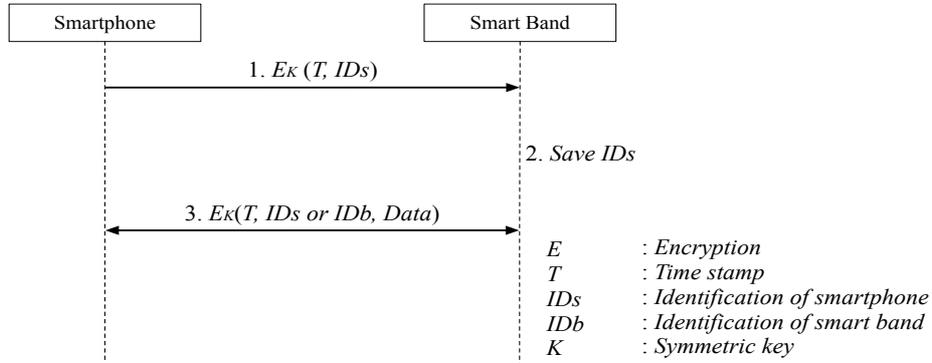

**Fig. 7.** Smartphone and smart band connection and communication protocol

### 4.3 Verification of Security Measures

We use Scyther to verify the security of the protocols mentioned above, and the results and codes are available in Github [25]. First, we verify the communication protocol between the smartphone and the web server. The process of exchanging the symmetric key using the station-to-station protocol is assumed to be safe and excluded from the verification. We checked the user ID (*ID*) the user password (*PW*) data transmitted between the smartphone and the web server (*PhoneData*, *ServerData*), the timestamp (*T*), and the symmetric key (*kir*) exchanged using the station-to-station protocol. As a result of verification of the protocol, all of them are secure and there is no attacker in the protocol. The Scyther verification code and result of smartphone and web server protocol is shown in **Fig. 8**. Second, we verify the communication and connection protocol between the smartphone and the smart band. The process of generating the symmetric key is assumed to be secure and is excluded from the verification Because it is used for the initial connection setting and the user reads the number and inputs it into the smartphone directly. We checked the smartphone connection request message (*connectionReq*), the smart band connection response message (*connectionRes*), data transmitted between the smartphone and the smart band (*PhoneData*, *BandData*), the timestamp (*T*), and the symmetric key (*kir*). As a result of verification of the protocol, all of them are secure and there is no attacker in the protocol. The Scyther verification code and result of the smartphone and smart band protocol is shown in **Fig. 9**.

```
1  /* Communication protocol between smartphone and web server*/
2
3  hashfunction hash;
4
5  usertype Key;
6  usertype Timestamp;
7  usertype SmartPhoneData;
8  usertype WebServerData;
9  usertype Password;
10 usertype Identification;
11
12 protocol pw(P,W)
13 {
14     role P
15     {
16         const PW: Password;
17         const ID: Identification;
18         const kir: Key;
19         fresh T1: Timestamp;
20         fresh T2: Timestamp;
21         fresh PhoneData: SmartPhoneData;
22         var T3: Timestamp;
23         var ServerData: WebServerData;
24
25         send_1 (P, W, {T1, ID, hash(PW)}pk(W) );
26         send_2 (P, W, {T2, P, PhoneData}kir );
27         recv_3 (W, P, {T3, W, ServerData}kir);
28
29         claim_p1 (P, Secret, PW);
30         claim_p2 (P, Secret, ID);
31         claim_p3 (P, Secret, PhoneData);
32         claim_p4 (P, Secret, ServerData);
33         claim_p5 (P, Secret, kir);
34         claim_p6 (P, Secret, T1);
35         claim_p7 (P, Secret, T2);
36     }
37
38     role W
39     {
40         const PW: Password;
41         const ID: Identification;
42         const kir: Key;
43         var T1: Timestamp;
44         var T2: Timestamp;
45         var PhoneData: SmartPhoneData;
46         fresh T3: Timestamp;
47         fresh ServerData: WebServerData;
48
49         recv_1 (P, W, {T1, ID, hash(PW)}pk(W) );
50         recv_2 (P, W, {T2, P, PhoneData}kir );
51         send_3 (W, P, {T3, W, ServerData}kir);
52
53         claim_w1 (W, Secret, PW);
54         claim_w2 (W, Secret, ID);
55         claim_w3 (W, Secret, PhoneData);
56         claim_w4 (W, Secret, ServerData);
57         claim_w5 (W, Secret, kir);
58         claim_w6 (W, Secret, T1);
59         claim_w7 (W, Secret, T2);
60     }
61 }
```

Scyther results : verify

| Claim | | | | Status | | Comments |
|-------|---|-------|------------------|--------|----------|------------|
| pw | P | pw,p1 | Secret PW | **Ok** | Verified | No attacks. |
| | | pw,p2 | Secret ID | **Ok** | Verified | No attacks. |
| | | pw,p3 | Secret PhoneData | **Ok** | Verified | No attacks. |
| | | pw,p4 | Secret ServerData | **Ok** | Verified | No attacks. |
| | | pw,p5 | Secret kir | **Ok** | Verified | No attacks. |
| | | pw,p6 | Secret T1 | **Ok** | Verified | No attacks. |
| | | pw,p7 | Secret T2 | **Ok** | Verified | No attacks. |
| | W | pw,w1 | Secret PW | **Ok** | Verified | No attacks. |
| | | pw,w2 | Secret ID | **Ok** | Verified | No attacks. |
| | | pw,w3 | Secret PhoneData | **Ok** | Verified | No attacks. |
| | | pw,w4 | Secret ServerData | **Ok** | Verified | No attacks. |
| | | pw,w5 | Secret kir | **Ok** | Verified | No attacks. |
| | | pw,w6 | Secret T1 | **Ok** | Verified | No attacks. |
| | | pw,w7 | Secret T2 | **Ok** | Verified | No attacks. |

Done.

**Fig. 8.** Scyther verification code and result of smartphone - web server protocol.

```
1 /*  Connection and Communication protocol between smartphone and smart band*/
2
3 hashfunction hash;
4
5 usertype Key;
6 usertype Timestamp;
7 usertype ConnectionRequest;
8 usertype ConnectionResponse;
9 usertype SmartPhoneData;
10 usertype SmartBandData;
11
12 protocol pb(P,B)
13 {
14    role P
15    {
16        const kir: Key;
17        fresh T1: Timestamp;
18        fresh T3: Timestamp;
19        fresh connectionReq: ConnectionRequest;
20        fresh PhoneData: SmartPhoneData;
21        var T2: Timestamp;
22        var T4: Timestamp;
23        var connectionRes: ConnectionResponse;
24        var BandData: SmartBandData;
25
26        send_1 (P, B, {T1, connectionReq}kir);
27        recv_2 (B, P, {T2, connectionRes}kir);
28        send_3 (P, B, {T3, P, PhoneData}kir);
29        recv_4 (B, P, {T4, B, BandData}kir);
30
31        claim_p1 (P, Secret, connectionReq);
32        claim_p2 (P, Secret, connectionRes);
33        claim_p3 (P, Secret, PhoneData);
34        claim_p4 (P, Secret, BandData);
35        claim_p5 (P, Secret, kir);
36        claim_p6 (P, Secret, T1);
37        claim_p7 (P, Secret, T2);
38        claim_p8 (P, Secret, T3);
39        claim_p9 (P, Secret, T4);
40    }
41
42    role B
43    {
44        const kir: Key;
45        var T1: Timestamp;
46        var T3: Timestamp;
47        var connectionReq: ConnectionRequest;
48        var PhoneData: SmartPhoneData;
49        fresh T2: Timestamp;
50        fresh T4: Timestamp;
51        fresh connectionRes: ConnectionResponse;
52        fresh BandData: SmartBandData;
53
54        recv_1 (P, B, {T1, connectionReq}kir);
55        send_2 (B, P, {T2, connectionRes}kir);
56        recv_3 (P, B, {T3, P, PhoneData}kir);
57        send_4 (B, P, {T4, B, BandData}kir);
58
59        claim_b1 (B, Secret, connectionReq);
60        claim_b2 (B, Secret, connectionRes);
61        claim_b3 (B, Secret, PhoneData);
62        claim_b4 (B, Secret, BandData);
63        claim_b5 (B, Secret, kir);
64        claim_b6 (B, Secret, T1);
65        claim_b7 (B, Secret, T2);
66        claim_b8 (B, Secret, T3);
67        claim_b9 (B, Secret, T4);
68    }
69 }
```

Scyther results : verify

| Claim | | | | Status | | Comments |
|-------|---|------|----------------------|----|----------|------------|
| pb | P | pb,p1 | Secret connectionReq | Ok | Verified | No attacks. |
| | | pb,p2 | Secret connectionRes | Ok | Verified | No attacks. |
| | | pb,p3 | Secret PhoneData | Ok | Verified | No attacks. |
| | | pb,p4 | Secret BandData | Ok | Verified | No attacks. |
| | | pb,p5 | Secret kir | Ok | Verified | No attacks. |
| | | pb,p6 | Secret T1 | Ok | Verified | No attacks. |
| | | pb,p7 | Secret T2 | Ok | Verified | No attacks. |
| | | pb,p8 | Secret T3 | Ok | Verified | No attacks. |
| | | pb,p9 | Secret T4 | Ok | Verified | No attacks. |
| | B | pb,b1 | Secret connectionReq | Ok | Verified | No attacks. |
| | | pb,b2 | Secret connectionRes | Ok | Verified | No attacks. |
| | | pb,b3 | Secret PhoneData | Ok | Verified | No attacks. |
| | | pb,b4 | Secret BandData | Ok | Verified | No attacks. |
| | | pb,b5 | Secret kir | Ok | Verified | No attacks. |
| | | pb,b6 | Secret T1 | Ok | Verified | No attacks. |
| | | pb,b7 | Secret T2 | Ok | Verified | No attacks. |
| | | pb,b8 | Secret T3 | Ok | Verified | No attacks. |
| | | pb,b9 | Secret T4 | Ok | Verified | No attacks. |

Done.

**Fig. 9.** Scyther verification code and result of smartphone - smart band protocol.

## 5    Conclusion

In this work, we derived vulnerabilities and security requirements for smart band system using threat modeling such as DFD, STRIDE, and Attack Tree. Vulnerabilities in the connection process of smart bands were found by analyzing the threats and attacks obtained through threat modeling. By exploiting these vulnerabilities, we could gain the system privilege easily. This shows that our threat modeling method is effective in detecting vulnerabilities of smart band system and security requirements analysis is useful for rebuilding a smart band system. We proposed security measures to respond to vulnerabilities in the connection process and other possible attacks. To validate the proposed security measures mathematically, we formalized the protocols and then verified that those measures are secure using Scyther. After establishing security measures against the threats and attacks derived from threat modeling, the system should be rebuilt considering the security measures. Applying the security measures to the system can result in new problems such as key management. Therefore, it is important to check the security after rebuilding a system by repeating the methodology present in this paper for the secure system.

## References

1. W. Zhou and S. Piramuthu, "Security/privacy of wearable fitness tracking IoT devices," in *Proc. 9th Iberian Conf. Inf. Syst. Technol.*, 2014, pp. 1–5.
2. M. LEE, K. Lee, J. Shim, S. J. Cho, and J. Choi, "Security threat on wearable services: empirical study using a commercial smartband," in *Consumer Electronics-Asia (ICCE-Asia), IEEE International Conference on*. IEEE, 2016. p. 1-5.
3. R. Goyal, N. Dragoni, and A. Spognardi, "Mind the tracker you wear: A security analysis of wearable health trackers," in *Proc. ACM Symp. Appl. Comput.*, 2016, pp. 131–136.
4. S. Seneviratne, Y. Hu, T. Nguyen, G. Lan, S. Khalifa, K. Thilakarathna, M. Hassan, and A. Seneviratne, "A survey of wearable devices and challenges," *IEEE Commun. Surveys Tuts.*, vol. 19, no. 4, pp. 2573–2620, 2017
5. NIST, Trustworthy Information System [Online]. Available: https://www.nist.gov/itl/trustworthy-information-systems. Accessed on: Dec 6, 2018.
6. H. F. Tipton, M. Krause, Information Security Management Handbook, Boca Raton, FL, USA:CRC Press, 2003.
7. A. Shostack, *Threat Modeling: Designing for Security*, 1st ed. John Wiley \& Sons Inc., Indianapolis, IN, 2014.
8. A. Shostack, "Experiences Threat Modeling at Microsoft", Modeling Security Workshop, Toulouse, 2008
9. STRIDE threat model of Microsoft. [Online]. Available: https://msdn.microsoft.com/en-us/library/ee823878 (v=cs.20).aspx. Accessed on: Dec 6, 2018.
10. I. Williams, X. Yuan, "Evaluating Effectiveness of Microsoft Treat Modeling Tool", *ISCD Conference*, 2015, October 2015.
11. B. Schneier, "Attack trees: Modeling security threats," *Dr. Dobb's Journal*, Dec 1999.
12. C. Cremers, "The Scyther Tool: Verification, falsification, and analysis of security protocols," in *Computer Aided Verification (CAV)*, ser. LNCS, vol. 5123. Springer, 2008, pp. 414–418.
13. M. Rahman, B. Carbunar, and M. Banik, "Fit and vulnerable: Attacks and defenses for a health monitoring device," *arXiv 1304.5672*, 2013
14. H. Fereidooni, T. Frassetto, M. Miettinen, A.-R. Sadeghi, and M. Conti. "Fitness Trackers: Fit for Health but Unfit for Security and Privacy," in *Connected Health: Applications, Systems and Engineering Technologies (CHASE), 2017 IEEE/ACM International Conference on*. IEEE, 2017. pp. 19-24
15. P. K. Akshay Dev and K. P. Jevitha, "STRIDE Based Analysis of the Chrome Browser Extensions API," in *Proceedings of the 5th International Conference on Frontiers in Intelligent Computing: Theory and Applications : FICTA 2016, Volume 2*, S. C. Satapathy, V. Bhateja, S. K. Udgata, and P. K. Pattnaik, Eds., ed Singapore: Springer Singapore, 2017, pp. 169-178.

16. A. Karahasanovic, P. Kleberger, M. Almgren, "Adapting Threat Modeling Methods for the Automotive Industry," [Online]. Available: http://publications.lib.chalmers.se/records/fulltext/252083/local_252083.pdf. Accessed on: Oct 24, 2018.

17. M. Cagnazzo, M. Hertlein, T. Holz, and N. Pohlmann "Threat Modeling for Mobile Health Systems," in *Wireless Communications and Networking Conference Workshops (WCNCW), 2018 IEEE*. IEEE, 2018. pp. 314-319

18. D. Basin, C. Cremers, and S. Meier, "Provably Repairing the ISO/IEC 9798 Standard for Entity Authentication," *Proc. 1st Int'l Conf. Principles of Security and Trust* (POST 12), LNCS 7215, P. Degano and J.D. Guttman, eds., 2012, pp. 129–148.

19. C. Cremers, "Key exchange in IPsec revisited: formal analysis of IKEv1 and IKEv2," *in European conference on research in computer security (ESORICS)*, Leuven, Belgium, Sep. 2011.

20. Microsoft, SDL Threat Modeling Tool. [Online]. Available: https://www.microsoft.com/en-us/sdl/adopt/threatmodeling.aspx. Accessed on: Dec 6, 2018.

21. B. Potter, "Microsoft SDL threat modelling tool," *Network Security*, vol. 2009, no. 1, pp. 15–18, 2009.

22. A. K. Das, P. H. Pathak, C.-N. Chuah, and P. Mohapatra, "Uncovering privacy leakage in BLE network traffic of wearable fitness trackers," in *Proc. 17th Int. Workshop Mobile Comput. Syst. Appl. (HotMobile)*, 2016, pp. 99–104.

23. Smart band attack demo video, [Online]. Available: https://youtu.be/QFb1AV7yUas. Accessed on: Dec 6, 2018.

24. W. Diffie, P. C. Van Oorschot, and M. J. Wiener, "Authentication and authenticated key exchanges," *Designs, Codes, Cryptography.*, vol. 2, no. 2, pp. 107–125, 1992

25. Scyther verification for security measures, [Online]. Available: *https://github.com/hausdorfff/Protocol-Verification*. Accessed on: Dec 6, 2018.